\documentclass[twocolumn]{aastex7}
\usepackage{amsmath}

\newcommand{\A}[1]{\textcolor{blue}{A{#1}}}
\newcommand{\B}[1]{\textcolor{red}{B{#1}}}

\begin{document}

\title{The Role of Reconnection at Magnetic Separators in Complex Solar Flare Ribbons}

\correspondingauthor{Graham Barnes}

\author[orcid=0000-0003-3571-8728,gname=Graham,sname=Barnes]{Graham Barnes}
\affiliation{NorthWest Research Associates}
\email[show]{graham@nwra.com}

\author[orcid=0000-0001-5661-9759,gname=Karin,sname=Dissauer]{Karin Dissauer}
\affiliation{NorthWest Research Associates}
\email{dissauer@nwra.com}
 
\begin{abstract}

Solar flare ribbons, manifesting as transient brightenings in the chromosphere, are believed to trace out the footpoints of magnetic field lines that are reconnecting higher in the solar atmosphere.
These field lines lie in a separatrix or quasi-separatrix layer that separates domains of different magnetic connectivity and hence forms a natural location for reconnection.
Solar flares are typically characterized as being circular ribbon flares, two-ribbon flares, or complex ribbon flares based on the number and shape of the ribbons. 
There are relatively well-developed models to explain the first two types of flares based on the location of the reconnection powering the flare.
The case of complex ribbons is less well understood, but is often posited to be a result of reconnection at multiple locations.
We demonstrate here that reconnection at a magnetic separator connecting two coronal null points can naturally explain the complex ribbons observed for two events, an M2.9 class flare from NOAA AR\,11112 (SOL2010-10-16T19:12), and an X2.2 class flare from NOAA AR\,11158  (SOL2011-02-15T01:56).

\end{abstract}

\keywords{\uat{Solar flares}{1496} --- \uat{Solar magnetic reconnection}{1504}}

\section{Introduction}\label{sec:intro}

The energy released in a solar flare is believed to be stored in the coronal magnetic field, and released by way of magnetic reconnection.  
Localized, enhanced emission, most often observed in H$\alpha$ and 1600\,\AA\ ultraviolet images and referred to as flare ribbons, occurs when highly energetic particles accelerated at the coronal reconnection site collide with lower, denser chromospheric material.
Magnetic field of different flux systems continually being drawn into the reconnection region leads to a ribbon's sequential brightenings and sometimes complex shape.
The magnetic field in the corona is rarely observed, so flare ribbons are a key to understanding the flaring process as they provide one of the observable signatures of the consequences of flare reconnection.

The morphology of flare ribbons is generally characterized as circular \citep[e.g.,][]{massonetal2009}, two-ribbon \citep[e.g.,][]{Fletcher2001}, or complex  \citep[e.g.,][]{chenetal2020}.
There does not appear to be a consensus on precise definitions for the different types of ribbons, so for the purposes of this paper, we define a circular ribbon as one consisting of a closed contour, two ribbons as being composed of two disjoint, open segments, one in each polarity of the magnetic field, and a complex ribbon as one composed of at least three disjoint, open segments with at least one segment in each polarity; a circular ribbon may also be accompanied by other ribbons that are not necessarily closed.

Circular ribbons are usually associated with reconnection at a magnetic null point in the corona \citep{massonetal2009}.
A magnetic null point is a location where the magnetic field vector vanishes, and is a natural location for reconnection to occur because domains of different magnetic connectivity lie in close proximity to a null.
The requirement that the field divergence vanish determines the structure of the field in the vicinity of the null: For a positive (negative) null, two spine field lines point towards (away from) the null, while a fan surface contains field lines pointing away from (towards) the null \citep{Longcope2005}.
The direction of the spine field lines (i.e., towards or away from the null) defines the type of null.
We shall refer to the intersection of the fan surface with the chromosphere as the fan trace.
The fan trace of an isolated null point typically forms a closed curve at the chromosphere.
This provides a natural explanation for a circular flare ribbon, which is expected to lie within a single magnetic polarity for a given event, and is often accompanied by a bright ``kernel'' at each spine footpoint \citep{PontinPriest2022}, one of which lies within the null separatrix surface.
A number of studies have found good correspondence between null fan traces, typically in potential field extrapolations, and circular ribbons observed in either H$\alpha$ data \citep[e.g,][]{WangLiu2012}, or in ultraviolet images \citep[e.g.,][]{UgarteUrraetal2007}.

Two-ribbon flares are often associated with the presence of a twisted magnetic flux rope. The CSHKP model \citep[see][for an overview]{Longcopeetal2007} provides an explanation in two dimensions for two-ribbon flares based on reconnection occurring at a current sheet that forms below an erupting flux rope.
This model was extended to the ``standard model in three dimensions'' \citep{AulanierJanvierSchmieder2012,Aulanieretal2013,Janvieretal2013,PontinPriest2022}, which includes the reconnection below the flux rope of the CSHKP model, but incorporates aspects that are not present in two dimensions.
In the simplest version of the standard model in 3D, the ribbons are J-shaped, with the straight parts of the ribbons approximately parallel.
This model has been used to interpret the ribbons in many events \citep[see][for an overview, and, e.g., \cite{Janvieretal2014,Savchevaetal2015,Savchevaetal2016,Zhaoetal2016}]{Dudiketal2025}.

Complex ribbon flares are often identified as those that do not easily fit into either the circular or two ribbon categories \citep[e.g.,][]{chenetal2020}, and are sometimes attributed to reconnection at more than one location.
A number of case studies have related the magnetic topology from an extrapolation to the locations of such flare ribbons. 
\cite{Joshietal2017} and \cite{Massonetal2017}, for two different cases, find a likely explanation for the event each considered to be a combination of null point reconnection resulting in a (quasi-) circular ribbon and reconnection from a flux rope lying within the null separatrix surface resulting in two inner ribbons.
\cite{Joshietal2022} identify what they characterize as a typical two ribbon configuration but with two additional, remote ribbons present.
They propose a similar explanation to the events studied by \cite{Joshietal2017,Massonetal2017}, with the inner two ribbons resulting from reconnection associated with a flux rope, but the remote ribbons being a result of reconnection between a flux rope and overlying field without a null point being present.

\begin{figure}
\plotone{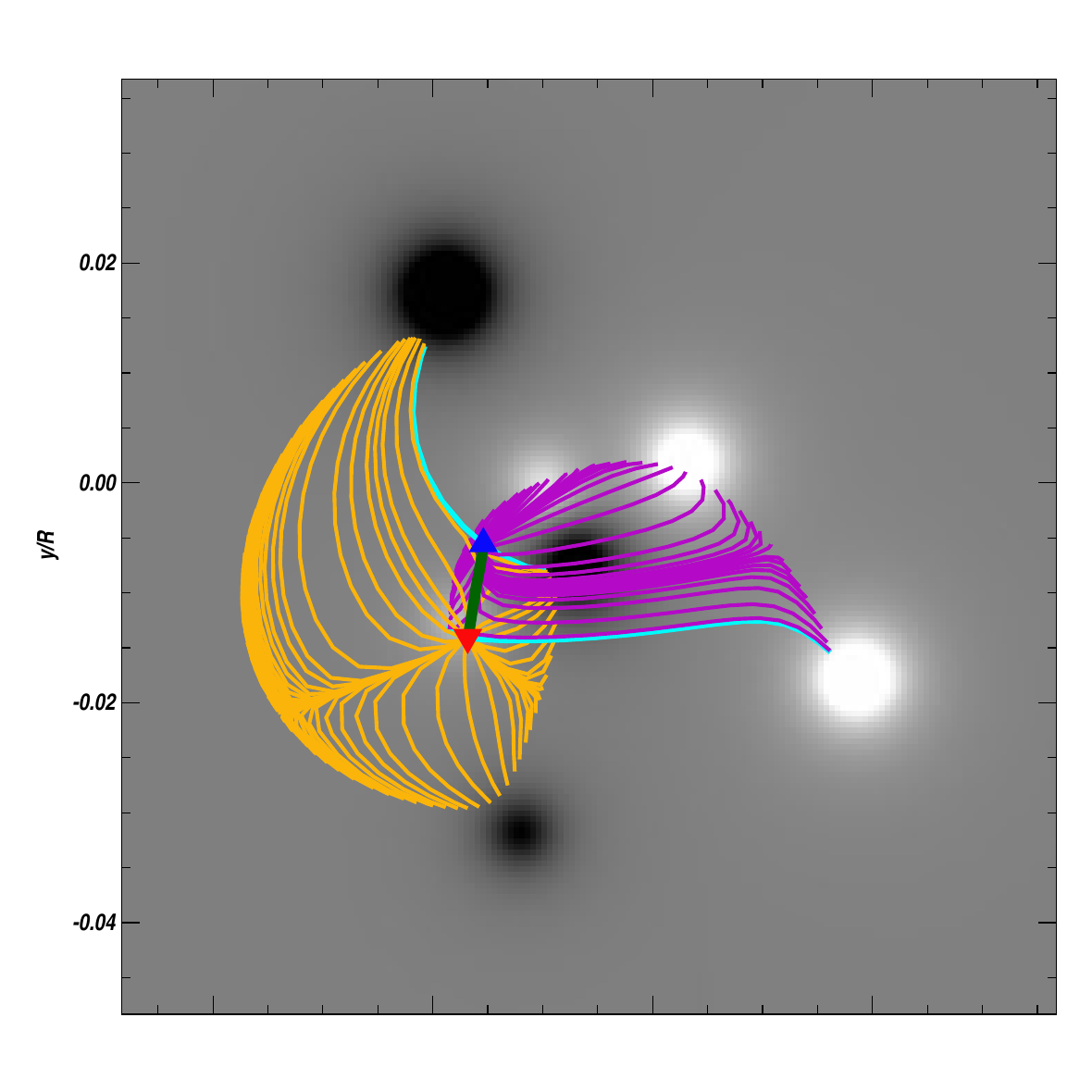}
\plotone{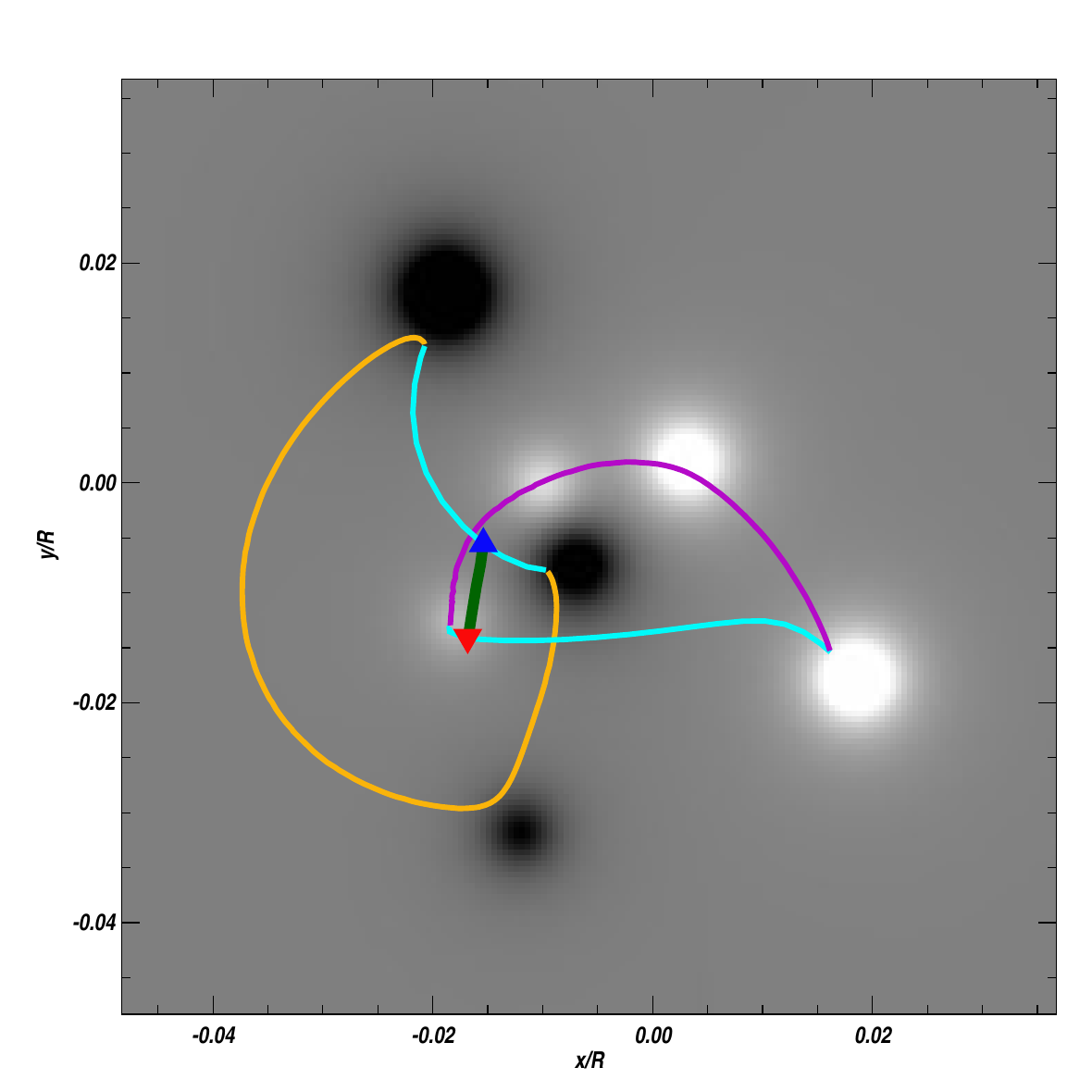}
\caption{The magnetic skeleton in a potential field model containing two opposite type coronal magnetic null points (red and blue triangles).
The background image is the radial component of the model field, and the spine field lines are shown as cyan curves.
In the top panel, field lines in the separatrix surfaces of the nulls are plotted in orange and magenta. 
A separator field line (green) connects the two nulls at the intersection of their fan surfaces.
In the bottom panel, the footpoints at the lower boundary of separatrix field lines are shown. 
The fan traces are open curves, with the end of each fan trace corresponding to the location of the intersection with the lower boundary of a spine field line from the opposite type null.
\label{fig:tworibbon}}
\end{figure}

A magnetic separator provides a natural location for an additional site of reconnection than can give rise to complex ribbons. 
A separator is the intersection of two separatrix surfaces, and hence lies among multiple domains of different connectivity. 
In the simplest case, when two null points of opposite type are present, they can be connected by a separator field line (see Figure~\ref{fig:tworibbon}).
In this case, the edge of the fan surface of each null follows the spine field lines of the other null;
\citet{Plattenetal2014} refer to this type of separatrix surface as a cave.
When the spine field lines are closed (i.e., when they end at the solar surface) the fan trace for each null is an open curve that runs between the footpoints of the spine field lines of the opposite type null. 

As noted by \cite{BungeyTitovPriest1996}, the topological skeleton can also contain separatrix surfaces associated with bald patches. 
A bald patch is a stretch of polarity inversion line along which the horizontal field is directed from negative polarity to positive polarity, corresponding to dipped magnetic field lines \citep{Seehafer1986}. 
Field lines that graze this part of the polarity inversion line lie in the bald patch separatrix surface. 
A bald patch separatrix surface intersecting a null separatrix surface also forms a separator.

When multiple separatrix surfaces intersect the separatrix surface of one null point, the result is a fan trace comprised of multiple disjoint segments since the null dome has multiple openings, each bounded by the spine fields lines of one of the nulls connected to it by a separator, or by a field line in a bald patch separatrix surface; 
\citet{Plattenetal2014} refer to this type of separatrix surface, with more than one opening, as a tunnel.
Reconnection at one separator from such a null point would be expected to produce multiple flare ribbons.
The fan trace of the tunnel separatrix surface would have disjoint segments equal in number to the number of openings in the dome.
In addition, at least one segment of fan trace from the null at the other end of the separator would lie in an area of opposite polarity field.

For two solar flares, we illustrate that separator reconnection successfully predicts the approximate location of flare ribbons that are not part of the standard model.
For the first event, the topological skeleton contains three null points, with two caves and one tunnel separatrix surface, while the second event has an even more complex topology.

\section{The Coronal Field Model}

To model the coronal magnetic field, we use a Potential Field Source Surface \citep[PFSS;][]{AltschulerNewkirk1969,Schattenetal1969} model consisting of a potential magnetic field whose radial component matches the input map of $B_r$, and that has no horizontal field at the source surface. 
The input map is taken from the Helioseismic and Magnetic Imager \citep[HMI;][]{Scherreretal2012,Hoeksemaetal2014} on board the Solar Dynamics Observatory \citep[\textit{SDO};][]{PesnellThompsonChamberlin2012}, specifically from the series {\tt hmi.B\_720s}.
The implementation of the PFSS model presented here uses a single magnetogram as a boundary condition, combined with an assumption of asymmetry to get the boundary condition in the far hemisphere and guarantee zero monopole moment.
Specifically, the field is assumed to satisfy $B_r(R,\theta,\pi-\phi)=-B_r(R,\theta,\phi)$, with the value of $B_r$ given by the observations on the front side of the Sun, in the longitude range $-\pi/2<\phi<\pi/2$.
The model uses the SHTOOLS \citep{shtools} package to compute spherical harmonics that are accurate up to a maximum degree of approximately 2800.
This implementation resolves small scale features that may be important in the local magnetic topology of an active region, but is not suitable for the global topology because of the lack of information about the far side.

Magnetic null points in the PFSS model are located by using a Markov Chain Monte Carlo method to initiate a Newton-Raphson method, discussed further in \S\ref{sec:nullfinding} and Appendix~\ref{app:nullfinding}.
The separatrix surface of each relevant null point is determined based on the approach of \cite{HaynesParnell2010}. 

\section{Flare Observations}

The flare ribbon analysis for both events is performed on a sequence of 1600\,\AA~ultraviolet images from the Atmospheric Imaging Assembly \citep[AIA;][]{lemenetal2012} onboard \textit{SDO}.
Full spatial (1.5" sampled at 0.6") and temporal resolution (24~s) image data is used and standard SSWIDL software is applied to calibrate the images (using \texttt{aia\_prep.pro}). The image closest to the reference time of the HMI vector magnetogram for each event is used as a reference against which all other images are tracked and corrected for differential rotation (using \texttt{drot\_map.pro}). 
Flare ribbons are detected following the approach of \cite{kazachenkoetal2017} using $c=8$ for the cutoff intensity $I_c$ and correcting for image saturation due to CCD blooming using $I_{sat}=4000$~counts/s and selecting neighbors within 3 to 30 pixels in the x- and y-directions. In addition, we remove artifacts around this initial set of saturated pixels (independent of a threshold intensity) by detecting pixels based on the median absolute deviation estimated from pre-flare images \citep[][]{Maybhate2008}.

To analyze the evolution of the ribbons in more detail we construct time and heat maps \citep[][]{Dissaueretal2025}. In time maps each flare ribbon pixel is assigned a value based on the first time of its detection with respect to the flare onset time. Heat maps represent masks where flare ribbon pixels that occur multiple times at the same location or are longer-lived have a higher value within these masks.

\section{Case Study \#1: NOAA AR\,11112}\label{sec:AR11112}

\begin{figure*}
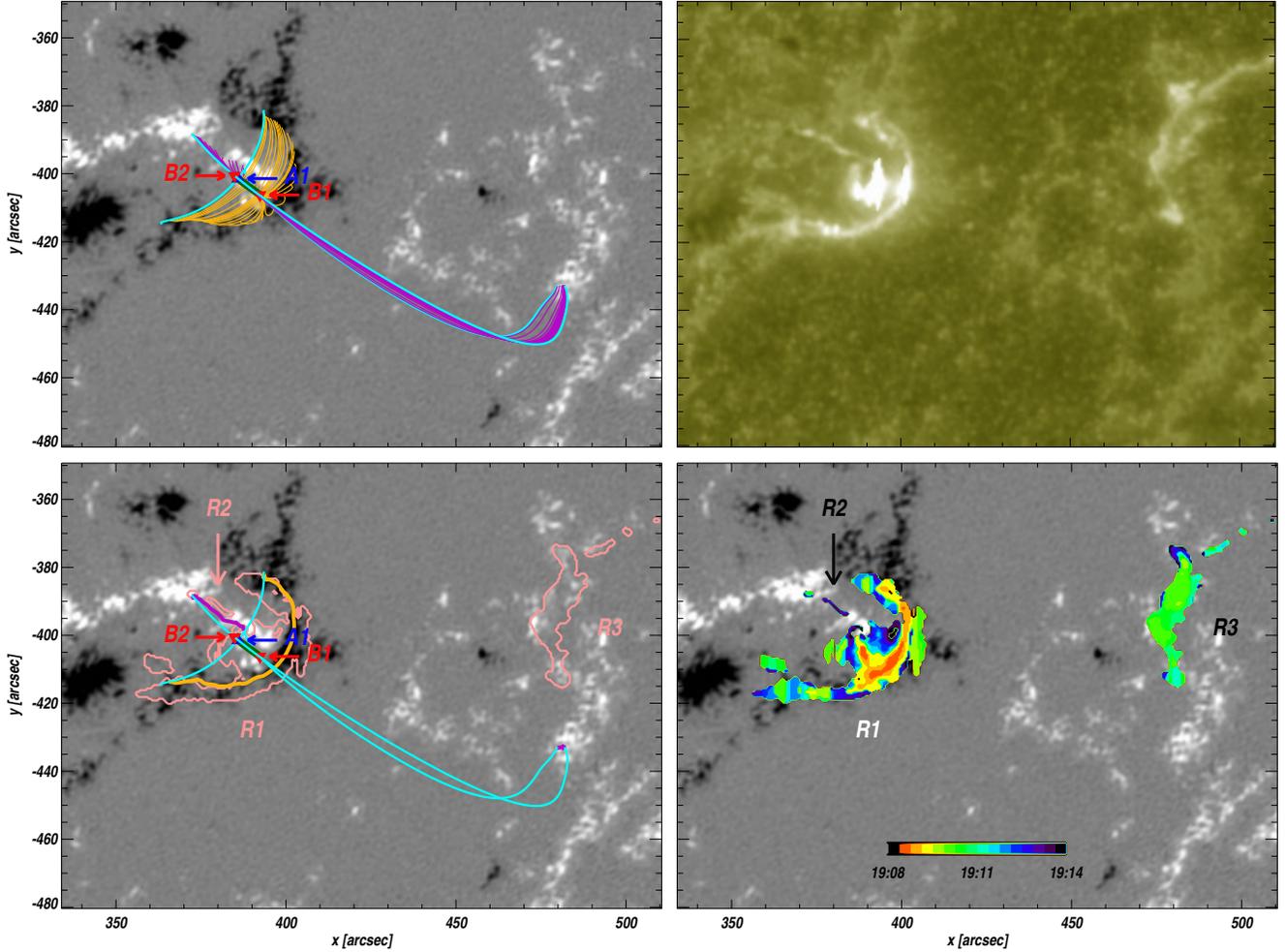

\gridline{\fig{2010-10-16T17-48-00_AR11112_Pr_nls_fan.eps}{0.57\textwidth}{}\hspace{-1.9cm}
	  \fig{2010-10-16T19-05-05_AR11112_1600.eps}{0.57\textwidth}{}}
\vspace{-1.8cm}
\gridline{\fig{2010-10-16T17-48-00_AR11112_Pr_nls_ftr_rbn.eps}{0.57\textwidth}{}\hspace{-1.9cm}
	  \fig{2010-10-16T17-48-00_AR11112_Pr_rbn.eps}{0.57\textwidth}{}}
\vspace{-0.8cm}
\caption{Elements of the magnetic skeleton in a PFSS model for NOAA AR\,11112 at 17:48\,TAI on 2010 October 16 (left) and flare ribbons for the M2.9 flare (right) on 2010-10-16.
Selected magnetic null points are shown as red and blue triangles, with color determined by null type, and their spine field lines (cyan curves).
The greyscale image is the radial component of the PFSS model, scaled to $\pm 10^3$\,G. 
In the top left panel, field lines in the separatrix surfaces of nulls \A{1} (magenta) and \B{1} (orange) are plotted, along with the separator field line (green) connecting the nulls (largely obscured by the spine field line above it, but clearly visible in Figure~\ref{fig:AR11112_skeleton_detail}) that delineates the intersection of the two separatrix surfaces.
In the bottom left panel, the footpoints at the lower boundary of separatrix field lines (fan traces) are shown, along with a salmon contour outlining the flare ribbons present in at least two frames of AIA data.
Gaps in the fan traces are spanned by spine field lines (e.g., the gap in the orange fan trace from null \B{1} is spanned by the spine field lines of null \A{1}).
The locations of the flare ribbons, visible in an AIA 1600\,\AA\ image at 19:05\,TAI in the upper right panel, are color coded by the time of first detection in the lower right panel. 
In the core of the active region, the locations of ribbon R1 and the fan trace of null \B{1} match reasonably well, as do the locations of ribbon R2 and one segment of the fan trace of null \A{1}, suggesting that reconnection at the separator connecting the nulls plays a role in this event.
Another segment of the fan trace of null \A{1} is also present in the vicinity of the remote ribbon R3, but the spatial agreement is worse than for the ribbons in the core of the region.
\label{fig:AR11112_skeleton}}
\end{figure*}

\begin{figure}
\plotone{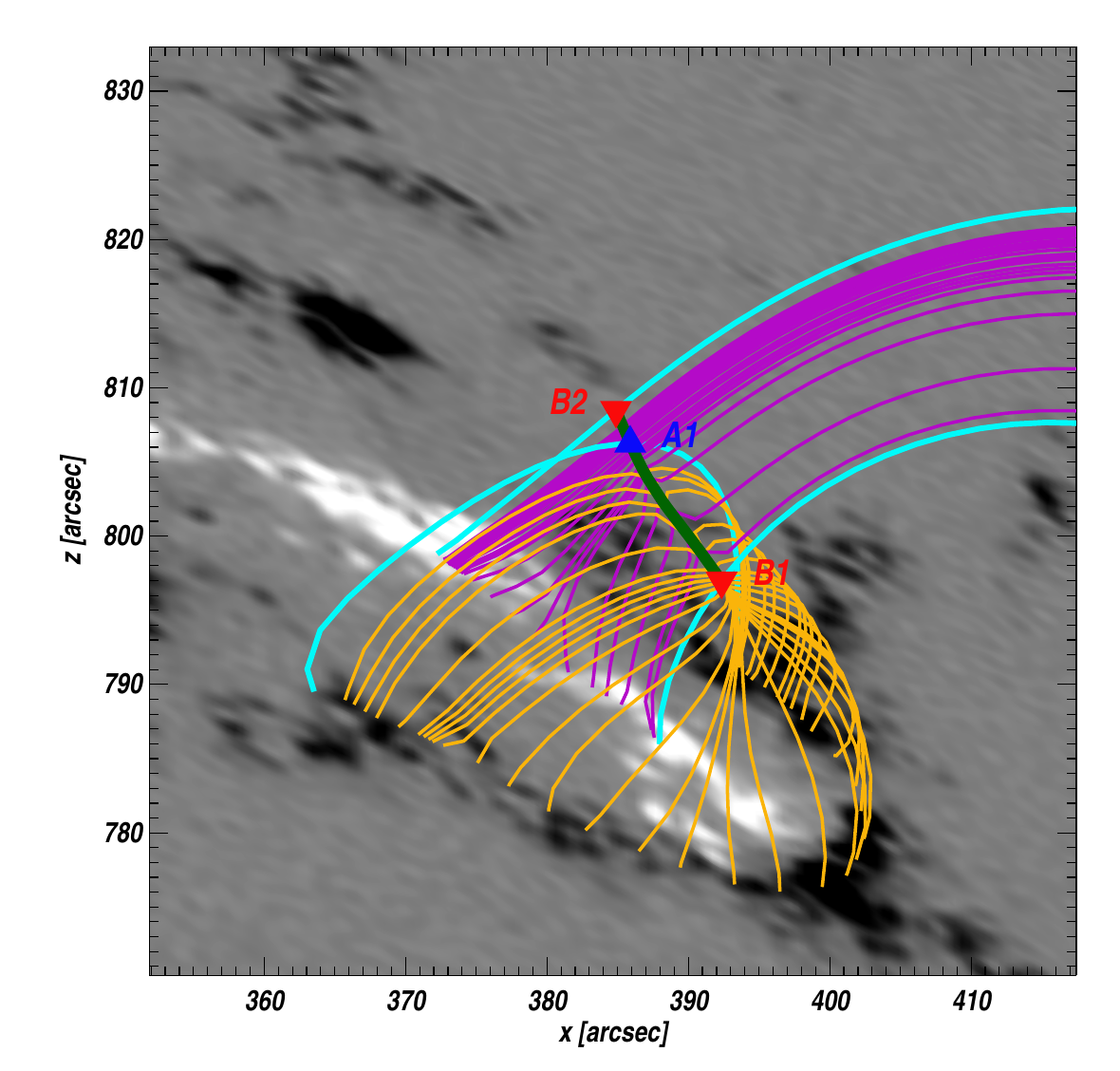}
\caption{The same elements of the magnetic skeleton shown in Figure~\ref{fig:AR11112_skeleton} for the core of NOAA AR\,11112 but viewed along the Sun's polar axis from the south.
The orange field lines lie in the fan surface of null \B{1}, which forms a cave, with the entrance to the cave following the spine field lines of null \A{1}.
The magenta field lines lie in the fan surface of null \A{1}, which forms a tunnel, with the two entrances to the tunnel following the spine field lines of nulls \B{1}, \B{2}.
Reconnection at the separator (green curve) connecting nulls \A{1} and \B{1} would result in a semi-circular flare ribbon along the fan trace of null \B{1}, and an approximately linear segment along one fan trace of null \A{1}.
\label{fig:AR11112_skeleton_detail}}
\end{figure}

To illustrate how separator reconnection could result in complex flare ribbons, we determined the magnetic skeleton above NOAA AR\,11112 on 2010 October 16 at 17:48\,TAI \citep{ReplicationData}, approximately an hour before the onset of an M2.9 class flare (SOL2010-10-16T19:12).
Figure~\ref{fig:AR11112_skeleton} shows the location of the flare ribbons as well as highlighting a few elements of the magnetic skeleton;
Figure~\ref{fig:AR11112_skeleton_detail} shows the same elements of the magnetic skeleton from a different vantage point. 
Three flare ribbons are identified: a semi-circular ribbon in negative polarity field labeled R1, a thin, linear ribbon in positive polarity field in the core of the active region labeled R2, and a remote ribbon R3 also in positive polarity field.
For clarity, the figures show only the separatrix surfaces associated with two selected null points (\A{1}, \B{1}), although a third null point (\B{2}) and its associated spine field lines are also shown.
The separatrix surface of null \A{1} intersects the separatrix surfaces of both nulls \B{1} and \B{2}, thus 
null \A{1} is connected by separator field lines to each of the opposite type nulls (\B{1}, \B{2}). 
The fan surface of null \A{1} is bounded on two sides by the spine field lines of nulls \B{1}, \B{2}, and by two segments of fan trace on the other sides.
Conversely, the fan surface of null \B{1} is bounded on one side by the spine field lines of null \A{1}, and by its fan trace on the other side.
Such a configuration could arise from a local separator bifurcation \citep{BrownPriest1999} in which an unstable second-order null point forms and immediately splits into a pair of nulls of opposite type connected by a separator, or a local double-separator bifurcation \citep{BrownPriest2001}, in which a single null point becomes an unstable third-order null that immediately splits into three null points, two of the same type as the original null and one of the opposite type.

The fan trace of null \B{1} (orange curve in the bottom left panel of Fig.~\ref{fig:AR11112_skeleton}) is roughly semicircular, with the two ends of the fan trace connected by the spine field lines of null \A{1}, as expected given the separator between them.
This fan trace approximately follows the semi-circular flare ribbon R1.

The fan trace of null \A{1} (magenta curves in the bottom left panel of Fig.~\ref{fig:AR11112_skeleton}) has two disjoint segments, with the ends of each segment connected to the other segment by the spine field lines of nulls \B{1}, \B{2}.
The segment of the fan trace of null \A{1} in the core of the active region is closely matched to the short segment of flare ribbon R2; the distant fan trace is in the vicinity of the remote flare ribbon R3, although somewhat displaced from it.
Thus reconnection at the separator connecting nulls \A{1} and \B{1}, combined with the presence of null \B{2}, naturally explains the presence of three flare ribbons observed for this event, and accurately predicts the location and shape of two of them.

This flare was used in a study of Extreme-Ultraviolet (EUV) Late Phase flares by \cite{chenetal2020}.
They identify three flare ribbons, which closely match the three we identify.
Although they use this event as an example of a circular ribbon flare, none of the ribbons are closed curves, which is why we consider it to be a complex ribbon event.
Based on computing the squashing factor $Q$ \citep{Titov2007}, they identify both spine-fan topology from a null point (our null \B{1}), and what they label as a plate-shaped quasi-separatrix layer (QSL).
In our analysis, the plate-shaped QSL is in fact a true separatrix surface associated with null \A{1}, and their ``dome--plate QSL'' is simply two null separatrix surfaces that intersect along the separator connecting the nulls, with the extent of the plate being limited by the presence of null \B{2}.

\section{Case Study \#2: NOAA AR\,11158}\label{sec:AR11158}

As a second example, we show results for the X2.2 flare from NOAA AR\,11158 on 2011 February 15 (SOL2011-02-15T01:56). 
This region, and multiple events that it gave rise to, have been very extensively studied.  
In many cases, the focus of the study was on the energy or helicity of the region \citep[e.g.,][]{Jingetal2012,Sunetal2012,TarrLongcopeMillhouse2013,TziotziouGeorgoulisLiu2013,Kazachenkoetal2015,Zhaoetal2016}, or on abrupt changes in the photospheric magnetic field \citep[e.g.,][]{Wangetal2012,Petrie2013}.  
However, several investigations focused on the topology of the region \citep[e.g.,][]{Sunetal2012b,Janvieretal2014,Zhaoetal2014}; 
the results of these studies are contrasted with our results below.

\begin{figure*}
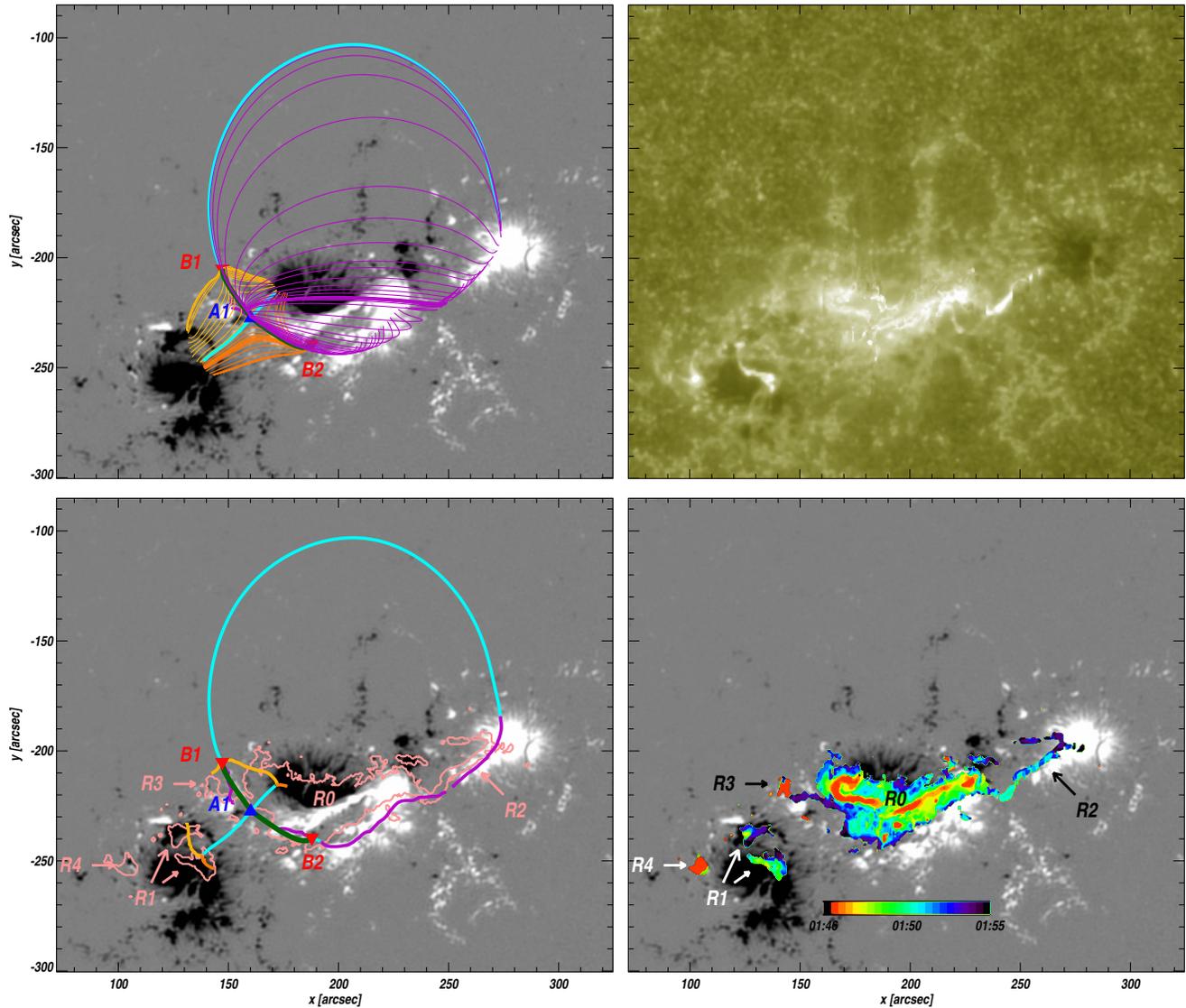

\gridline{\fig{2011-02-15T01-00-00_AR11158_Pr_nls_fan.eps}{0.53\textwidth}{}\hspace{-1.3cm}
          \fig{2011-02-15T01-51-29_AR11158_1600.eps}{0.53\textwidth}{}}
\vspace{-1.8cm}
\gridline{\fig{2011-02-15T01-00-00_AR11158_Pr_nls_ftr_rbn.eps}{0.53\textwidth}{}\hspace{-1.3cm}
          \fig{2011-02-15T01-00-00_AR11158_Pr_rbn.eps}{0.53\textwidth}{}}
\vspace{-0.8cm}
\caption{Elements of the magnetic skeleton in a PFSS model for NOAA AR\,11158 at 01:00\,TAI on 2011 February 15 (left) and flare ribbons for the X2.2 flare (right) on 2011-02-15 in the same format as Fig.~\ref{fig:AR11112_skeleton}.
In the top left panel, field lines in the separatrix surfaces of three nulls are plotted (two shades of orange and magenta), along with the separator field lines (green) connecting the nulls.  
In the bottom left panel, parts of the fan traces of nulls \B{1}, \B{2}, and \A{1} are in close proximity to the ribbons labeled R1, R2, and R3.
All the flare ribbons are visible in an AIA 1600\,\AA\ image at 01:51\,TAI shown in the upper right panel.
In the bottom right panel, the earliest appearance of the flare ribbons labeled R0 is in the J-shape often associated with a twisted magnetic flux rope. 
\label{fig:AR11158_skeleton}}
\end{figure*}

\begin{figure}
\plotone{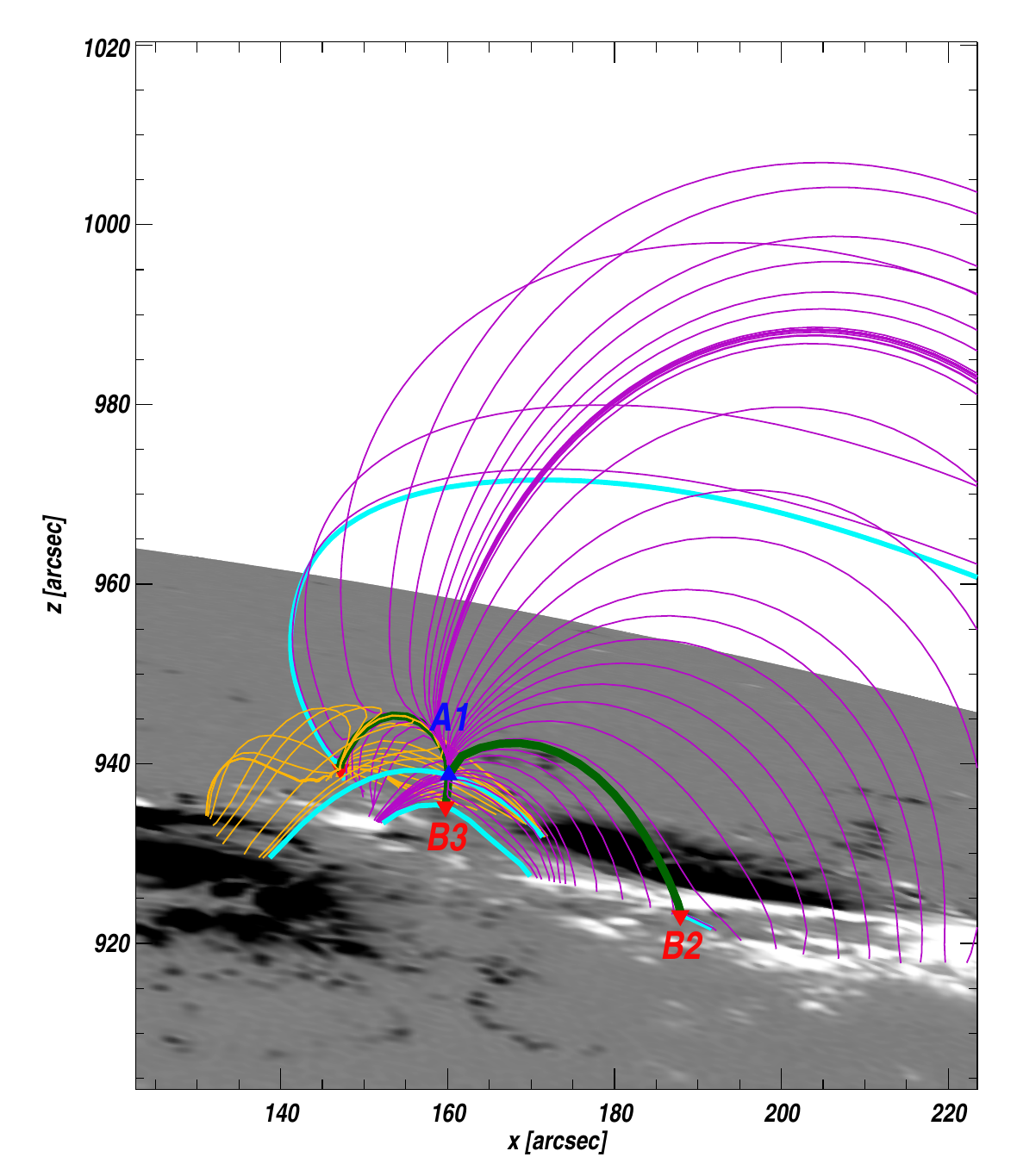}
\caption{Elements of the magnetic skeleton in the PFSS model for NOAA AR\,11158 but viewed along the Sun's polar axis from the south.
Field lines in separatrix surfaces of only nulls \A{1} and \B{1} are shown. 
Null \B{1} lies on the far side of its separatrix surface from this direction and thus is mostly obscured. 
An additional null point, \B{3}, is shown in this plot that was not included in Figure~\ref{fig:AR11158_skeleton} lying below null \A{1}, and connected to it by a separator.
This results in an additional gap in the fan trace of null \A{1}, not easily visible in Figure~\ref{fig:AR11158_skeleton}.
\label{fig:AR11158_skeleton_detail}}
\end{figure}

Figures~\ref{fig:AR11158_skeleton} and \ref{fig:AR11158_skeleton_detail} highlight a few pieces of the local topological skeleton as well as showing the location of the flare ribbons.  
In the earliest stages of the flare ribbons, the J-shaped structure predicted by the 3D standard model is visible in the lower right panel of Figure~\ref{fig:AR11158_skeleton} as the areas of ribbon R0 shaded yellow to red. 
Indeed, \cite{Janvieretal2014} use this event as an illustration of the 3D standard model.
At later times, multiple flare ribbons are also present outside the core of the active region, labeled R1, R2, and R3.
These ribbons are not predicted as part of the standard model, and make this a complex ribbon flare.

This region exhibits an extremely complex magnetic topology, with many null points and separators present in the field of view. 
For clarity, the left column of Figure~\ref{fig:AR11158_skeleton} shows only elements of the topological skeleton associated with three selected null points; 
Further details of the topological skeleton are given in Appendix~\ref{sec:AR11158_topology}.
These particular nulls are shown because of the reasonable qualitative agreement between their fan traces and the flare ribbons that appear later in the event. 

Null \A{1} is connected by separator field lines to three B-type nulls (with \B{3} only visible in Fig.~\ref{fig:AR11158_skeleton_detail}). 
Its fan trace has three disjoint segments, with the segments connected by way of arcs of spine field lines in the corona.
Reconnection at the separators connecting \A{1} with \B{1} and \B{2} would be expected to result in brightening at the chromosphere along the fan traces of all three nulls. 
The location of segments of the fan trace of \A{1} qualitatively agrees well with the location of ribbon R2 and R3.
Parts of the fan traces of nulls \B{1} and \B{2} qualitatively agree well with the ribbon R1.
Indeed, although we label this as a single ribbon, it consists of two distinct areas in the timing map, each corresponding to the location of a segment of fan trace from one of the two nulls.

The evolution of the flare ribbons is thus consistent with an event initiated by destabilizing a flux rope, as in the 3D standard model, which subsequently results in reconnection along two separators. 
Whether this exactly describes the process or not, we find that the local topological skeleton includes null point separatrices whose photospheric footpoints are in close proximity to the flare ribbons in the later stages of the event.

The predictions from separator reconnection are not perfect.
There is a small flare ribbon, R4, that appears early in the flare, and does not correspond to the fan trace of any null that we located.
Conversely, one segment of the fan trace of null \B{1} does not correspond to any flare ribbon that was identified in our analysis.
This fan trace is in the vicinity of one footpoint of the fluxrope, thus the potential field approximation is likely poor in this area, which may account for the lack of agreement.
Nonetheless, the agreement with other parts of the flare ribbons strongly suggests that separator reconnection played a role in this event.

Other investigations have generally found at most one coronal null point above this region. 
\cite{Sunetal2012b} found a single null point in a sequence of nonlinear force-free field (NLFFF) extrapolations, although their analysis ends prior to the time we consider so no direct comparison is possible.
Based on the location of the null and its spine field lines, it most closely corresponds to the null we label \B{1}. However, given the evolution of the photospheric magnetic field, it is possible that the topology has changed in the intervening time.

In a Magnetic Charge Topology model \citep[MCT;][]{BaumBratenahl1980}, \cite{TarrLongcopeMillhouse2013} find one A-type coronal null point whose location and spine field lines correspond very closely to our null \A{1}.
Interestingly, they also find B-type \textit{photospheric} null points in close proximity to each of the nulls we consider, each with a separator field line to the coronal null point. 
While certainly not true for all nulls in the MCT model, we speculate that in some cases, reducing the boundary to point sources causes coronal null points to become photospheric nulls.

\cite{Zhaoetal2014} produced a sequence of $Q$ maps for NLFFF extrapolations of this region.
They identify one null point, likely the one we label \A{1}\footnote{The NLFFF extrapolations used a cylindrical equal area grid, so it is difficult to exactly compare locations.}, and highlight several QSLs.
Their first QSL, labeled Q0, occurs in close proximity to the polarity inversion line between the flare ribbons, and corresponds to the flux rope in the 3D standard model. 
The flare ribbons visible in the earliest stages of the flare occur in close proximity to the photospheric footprint of this QSL.

\cite{Zhaoetal2014} identify two other main QSLs, labeled D1 and D2 in their Figure 3.
The location of QSL D1 corresponds very closely to the fan trace of our null point \B{1}, with two disjoint segments, while the location of QSL D2 corresponds closely to one segment of the fan trace of our null point \A{1}.
Although it is not labeled in their figure, there is also a QSL in close proximity to a second segment of the fan trace of our null \A{1}.
They further identify a hyperbolic flux tube \citep[HFT;][]{TitovHornigDemoulin2002} occurring at the intersection between these two QSLs that corresponds to the separator connecting our null points.
Thus we argue that these two main QSLs in their model for this region are in fact true separatrix surfaces, and that the morphology of the photospheric footprint of these QSLs is explained by the presence of additional elements of the magnetic skeleton (see Appendix~\ref{sec:AR11158_topology}).

In their illustration of the 3D standard model, \cite{Janvieretal2014} also consider the topology of this region. 
However, they mostly restrict their analysis to an area from $[x, y] = [170\arcsec, -240\arcsec]$ to $[240\arcsec, -210\arcsec]$, which does not include the flare ribbons that occur at later times and match with the fan traces in our analysis, so it is not possible to compare with our results.

\section{The challenges of locating magnetic null points}\label{sec:nullfinding}

In both the case studies presented here, true topological elements are found that have corresponding elements in the quasi-topology of other investigations.
Why were the true topological elements not identified?
There are multiple possible explanations, including spherical versus planar geometry and NLFFF versus potential field for the extrapolation, as well as preprocessing of the lower boundary. 
However, we hypothesize that interpolation on a grid of insufficient spatial resolution may be responsible.
In particular, the null finding algorithms of both \cite{HaynesParnell2007} and \cite{Nayaketal2020} rely on a trilinear approximation being valid within a grid element.
When two opposite type nulls lie within the same grid element, these methods will fail to identify them, and even if a grid element contains only one null, they can still fail if the variation of the field within the grid cell deviates sufficiently from the trilinear approximation.
The case of opposite type nulls present in the same grid element can be thought of in terms of the topological degree of the grid element. 
This case has the same topological degree as a grid element containing no null points, and thus these two scenarios cannot be distinguished based on the field on the surface of the grid element alone.
Thus even though the null points may be \textit{present} in the modeled field, they may not be \textit{located} because of the limits of the interpolation.

For the 2010 October 16 case, \cite{chenetal2020} apply the \cite{HaynesParnell2007} null finding algorithm to a potential field computed using Fourier transforms on a ``preprocessed'' boundary condition. 
They do not specify the grid on which the potential field was computed, or the extent to which the preprocessing modified the boundary. 
They locate one null point that corresponds to null \B{1} in our Figures~\ref{fig:AR11112_skeleton}, \ref{fig:AR11112_skeleton_detail}. 
Table~\ref{tbl:null_properties} gives the height and distance to the closest neighboring null point for each of the nulls shown in Figures~\ref{fig:AR11112_skeleton}, \ref{fig:AR11158_skeleton}.
The null located by \cite{chenetal2020} is $\sim 9$\,Mm from its closest neighbor while the additional nulls that we locate are a pair of opposite type in close proximity to each other (1.9\,Mm apart), consistent with our hypothesis. 

For the 2011 February 15 case, \cite{Zhaoetal2014} compute a NLFFF model on a grid of $178\times134\times134$, with pixel size $\sim$ 1.1\,Mm and focus on the volume above 2.2\,Mm. 
They locate one null point that likely corresponds to null \A{1} in our Figures~\ref{fig:AR11158_skeleton}, \ref{fig:AR11158_skeleton_detail}, but do not specify how the null point was located beyond commenting on a drastic change in direction of field lines in the neighborhood of the null, so it is not possible to check our hypothesis. 
However, the null they do locate is $\sim 3$\,Mm from its closest neighbor and at a height $\sim 5$\,Mm, while the other nulls we locate lie below 1\,Mm (see Table~\ref{tbl:null_properties}) and so any interpolation on a grid of this scale seems likely to miss some null points.

As previously noted, the study of \cite{Sunetal2012b} is not directly comparable because their analysis ends before the time we analyzed, but contains a potentially relevant statement. 
They used a NLFFF extrapolation on a Cartesian grid of dimensions $300 \times 300 \times 256$, with a 720\,km resolution, and applied the method of \cite{HaynesParnell2007} to locate null points. 
They note in an appendix that ``in some frames in the time series, multiple null points appear in adjacent computational cells'', and that ``both positive and negative nulls exist in the sample''. 
This is suggestive of null points being missed in some frames when nulls of opposite type fall within the same computational cell, or even are just in close enough proximity that the trilinear approximation does not accurately represent the field.

\begin{deluxetable}{lllll|lllll}
\tablewidth{0pt}
\tablecaption{Null heights and distance to closest null.}
\label{tbl:null_properties}
\tablehead{
\colhead{} & \multicolumn{3}{c}{AR\,11112} & \colhead{} & \colhead{} & \multicolumn{4}{c}{AR\,11112} \\
\cline{2-4}
\cline{7-10}
\colhead{Null} & \colhead{\A{1}} & \colhead{\B{1}} & \colhead{\B{2}} &
\colhead{} & \colhead{} &
\colhead{\A{1}} & \colhead{\B{1}} & \colhead{\B{2}} & \colhead{\B{3}}
}
\startdata
height [Mm] & 12.2         & 10.0         & 13.0         & \ & \ 
 & 5.3         &   0.7         & 0.2         & 2.8 \\
dist [Mm]   &  1.9 (\B{2}) &  8.9 (\A{1}) &  1.9 (\A{1}) &   & 
 & 2.6 (\B{3}) &  18.6 (\A{1}) & 0.9 (\A{2}) & 1.9 (\A{3}) \\
\enddata
\end{deluxetable}

Our null finding algorithm, described in greater detail in Appendix~\ref{app:nullfinding}, begins with a Markov chain that identifies volumes of low field strength, and refines the location of null points with a Newton-Raphson gradient based method initiated from points in the chain with field strength below a specified value.
The magnetic field and its derivatives are evaluated using the full spherical harmonic expansion without any approximation beyond truncating the expansion. 
While the field evaluation in the Newton-Raphson part of the computation is much more computationally intensive than the method of \cite{HaynesParnell2007}, this approach locates null points that are missed in the trilinear approximation, including opposite type nulls that are very close together.

\section{Summary}

Solar flares are typically characterized as being circular ribbon flares, two-ribbon flares, or complex ribbon flares.
Models for the first two morphologies of flare ribbons are relatively well developed. 
Circular ribbons are believed to result from reconnection at a coronal magnetic null point, whose separatrix surface forms a closed curve at the chromosphere, where the ribbons are observed.
Two-ribbon flares are described by the 3D standard model as resulting from reconnection at a current sheet below a flux rope.
The case of complex ribbons is less well understood, but is often posited to be a result of reconnection at multiple locations.
We have demonstrated here how some cases of complex ribbons can naturally be explained by reconnection at a magnetic separator connecting two coronal null points.

When null points are connected by a separator, the intersection with the chromosphere of the separatrix surface of each null must have at least one gap.
In the first case presented here, a semi-circular fan trace is produced when three null points are present, with a central null point of one type connected by separators to opposite type nulls on either side.
Such a configuration of nulls would result from a local double-separator bifurcation \citep{BrownPriest2001}.
Reconnection at one of the separators would naturally give rise to a semi-circular flare ribbon plus two linear flare ribbons, similar to what is observed for this event.

In the second case shown, the topological skeleton is much more complex, with multiple separators connecting pairs of null points.
The separatrix surface for each of the nulls considered is a tunnel with two or three openings, leading to multiple disjoint segments of fan trace for each null.
Reconnection at two of the separators associated with one null would be expected to produce flare ribbons at locations that are in close spatial proximity to those observed later in the flaring process.
The event is thus consistent with an initial destabilization of a flux rope, as in the 3D standard model, that triggers reconnection at the separators, resulting in the complex ribbons.

In both cases considered, other investigations found QSLs in close spatial proximity to the location of the true separatrix surfaces identified here.
We hypothesize that interpolating on too coarse a spatial grid may be the reason that some null points were not identified in other investigations. 
In particular, null points of opposite type that are separated by less than the size of a grid element are likely to be missed when interpolating.
Our method, which starts from volumes of low field strength determined by a Markov chain and evaluates the field at any point by summing the spherical harmonic expansion, is not subject to the same limitations.
We are thus able to infer the presence of magnetic separators that may have been missed in other investigations.

We emphasize separator reconnection here because the cases presented show evidence of flare ribbons close to the fan traces of pairs of nulls connected by a separator.
However, reconnection at a null point \citep[spine-fan or possibly torsional spine or torsional fan reconnection;][]{PriestPontin2009} with at least one separator would also give rise to an open (semi-circular) flare ribbon with bright kernels at the spine footpoints.

Multiple investigations consider cases described as having ``incomplete fan-spine configurations'', or ``semi-circular'' or ``quasi-circular'' ribbons, including C-shaped ribbons and arcs \citep[e.g.,][]{Politoetal2017,Suetal2018,MitraJoshi2021,MitraVeronigJoshi2023}.
In addition, \cite{Heetal2020} find that approximately half (35 of 71) two ribbon flares of class M1.0 and above in their sample are J-shaped, while the other half are not.
As seen in Figure~\ref{fig:tworibbon}, C-shaped pairs of ribbons are the expected outcome from reconnection at a single separator connecting two null points, while Figure~\ref{fig:AR11112_skeleton} illustrates that semi-circular or quasi-circular ribbons can be explained by separator reconnection when more than two nulls are connected by separators; null point reconnection when the null is connected by a separator to another null may provide another mechanism.
It thus seems likely that reconnection at magnetic separators plays an important role in many events that are not fully described by the standard flare model in three dimensions.
If our hypothesis about lack of resolution being responsible for null points not being located is correct, then it is also possible that null point reconnection is responsible for circular ribbon flares for which no null was located in prior investigations \citep[e.g.,][]{Liuetal2015,Haoetal2017}.

\begin{acknowledgments}
The material presented here is based upon work supported by NASA grants 80NSSC19K0087 and 80NSSC21K0738, and by NSF award 2154653 to NorthWest Research Associates.
\end{acknowledgments}

\begin{contribution}
GB was responsible for the topological analysis and for writing and submitting the manuscript.
KD was responsible for the flare ribbon analysis, and editing the manuscript.
\end{contribution}

\facilities{SDO(HMI and AIA)}

\software{SHTOOLS\citep{shtools}}

\appendix

\section{The Null-Finding algorithm}\label{app:nullfinding}

To locate magnetic null points, a Markov chain was used to determine volumes of low field strength, then a Newton-Raphson algorithm was used to refine the location of the null points.
Each chain was initiated at a random location in the volume. 
A new trial point was generated by taking a step in a random direction with a step size drawn from a Gaussian distribution with width $w=3\times10^{-4}\,R_\odot$.
The trial point was automatically accepted if the field strength decreased.
If the field strength increased, the trial point was accepted with a probability proportional to the ratio of the field strength times a prior probability at the present point versus at the trial point (the Metropolis-Hastings algorithm).
The column density of null points at a height $z$ above quiet Sun was estimated by \cite{LongcopeParnell2009} to be $N_n(z) \sim (z + d)^{-2}$ with $d=1.6$\,Mm. 
Thus the prior probability was chosen to be 
\begin{equation}
p(r) = \begin{cases} 
\bigg (\frac{r/R_\odot - 0.998}{0.002} \bigg )^{-3}, & r \ge R_\odot\\
1, & r < R_\odot
\end{cases}
\
\end{equation}
to be roughly consistent with the result of \cite{LongcopeParnell2009}.
This choice of prior is likely not optimal, but it was found that the chains converged much more rapidly than when a uniform prior was used. 
Likewise, the width of proposal distribution is likely not optimal, but it was found that the chains converged best for a value in the range $10^{-4}\,R_\odot \la w \la 10^{-3}\,R_\odot$.

\begin{figure*}
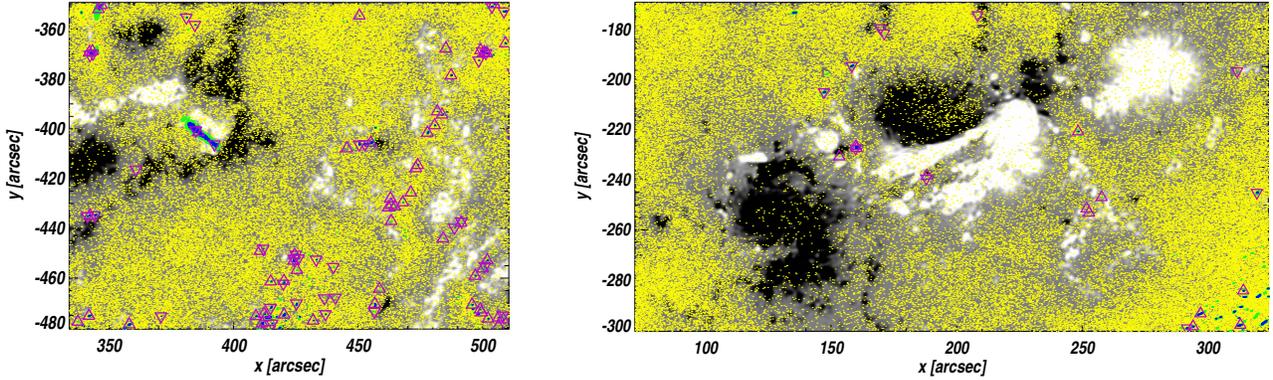

\gridline{\fig{2010-10-16T17-48-00_AR11112_Pr_mcmc_pts_nls.eps}{0.416\textwidth}{}\hspace{-0.0cm}
          \fig{2011-02-15T01-00-00_AR11158_Pr_mcmc_pts_nls.eps}{0.584\textwidth}{}}
\caption{Points in a Markov chain and null points in a PFSS model for NOAA AR\,11112 at 17:48\,TAI on 2010 October 16 (left) and NOAA AR\,11158 at 01:00\,TAI on 2011 February 15 (right).
Every $150^{\rm th}$ point in the chain is plotted in yellow.
All points in the chain with $B<10$\,G (green), and $B<5$\,G (blue) are plotted, along with the location of magnetic null points (magenta triangles).
\label{fig:null_finding}}
\end{figure*}

The Markov chain was constructed for a volume with $0.999\,R_\odot < r < 1.05\,R_\odot$.
Because many of the nulls lie just above $r=R_\odot$, a small volume below that surface ($\sim 3 \times$ the width of the proposal distribution) was included so the chain fully explores the volume just above $r=R_\odot$.
For AR\,11112, the horizontal extent of the chain is bounded by $21.87^\circ < \phi < 37.68^\circ$ in longitude and $-29.91^\circ < \theta < -21.28^\circ$ in latitude, which includes the full field of view shown in Figure~\ref{fig:AR11112_skeleton}.
For AR\,11158, the horizontal extent of the chain is bounded by $4.33^\circ < \phi < 20.51^\circ$ in longitude and $-17.99^\circ < \theta < -10.05^\circ$ in latitude, which includes the full east-west extent of the field of view shown in Figure~\ref{fig:AR11158_skeleton}, but does not extend as far to the north.
The number of points in each chain is $2\times10^7$.
If the same number of points were used on a uniform grid, the grid spacing would be $\sim 400$\,km.
However, as can be seen in Figure~\ref{fig:null_finding}, the Markov chain concentrates points above weak, mixed polarity radial field, and largely avoids the volume above areas of strong (unipolar) radial field.
Thus it more thoroughly explores volumes where null points are more likely to be found.

Points with a field strength $B<10$\,G are used to initiate the Newton-Raphson method. 
The result of the Newton-Raphson method is considered to be a null point (not just a local minimum in the field strength) if the field strength at the final iteration is $B < 10^{-8}$\,G, making it highly likely that a null point has been located.
Each null point is typically found multiple times because points in the Markov chain cluster in the vicinity of null points. 
Two null point candidates separated by a distance $d < 10^{-6}\,R_\odot$ are considered to be the same null. 
In practice, the distance is typically $d \la 10^{-10}\,R_\odot$, indicating that the location of each null is determined very precisely.
Figure~\ref{fig:null_finding} also shows the locations of all the null points located. 
For NOAA AR\,11112 in particular, there are many null points located above areas of weak, mixed polarity vertical field at the photosphere. 

For each region, seven chains with different sequences of random numbers were computed, and null points located.
For AR\,11112, a total of 94 distinct null points were located within the field of view shown. 
Of these, 87 were located from every chain, and the chain shown in Figure~\ref{fig:null_finding} found 93 null points; between 90 and 93 null points were located from each of the other chains.
For AR\,11158, a total of 25 distinct null points were located. 
Of these, 17 were located from every chain, and the chain shown in Figure~\ref{fig:null_finding} found 22 null points; between 20 and 22 null points were located from each of the other chains.
Clearly, no single chain has fully converged, since not all of the null points were located from any one chain. 
However, the large fraction of the total number of nulls found by every chain suggests that most if not all the nulls have been located from at least one chain.

\begin{figure*}
\plottwo{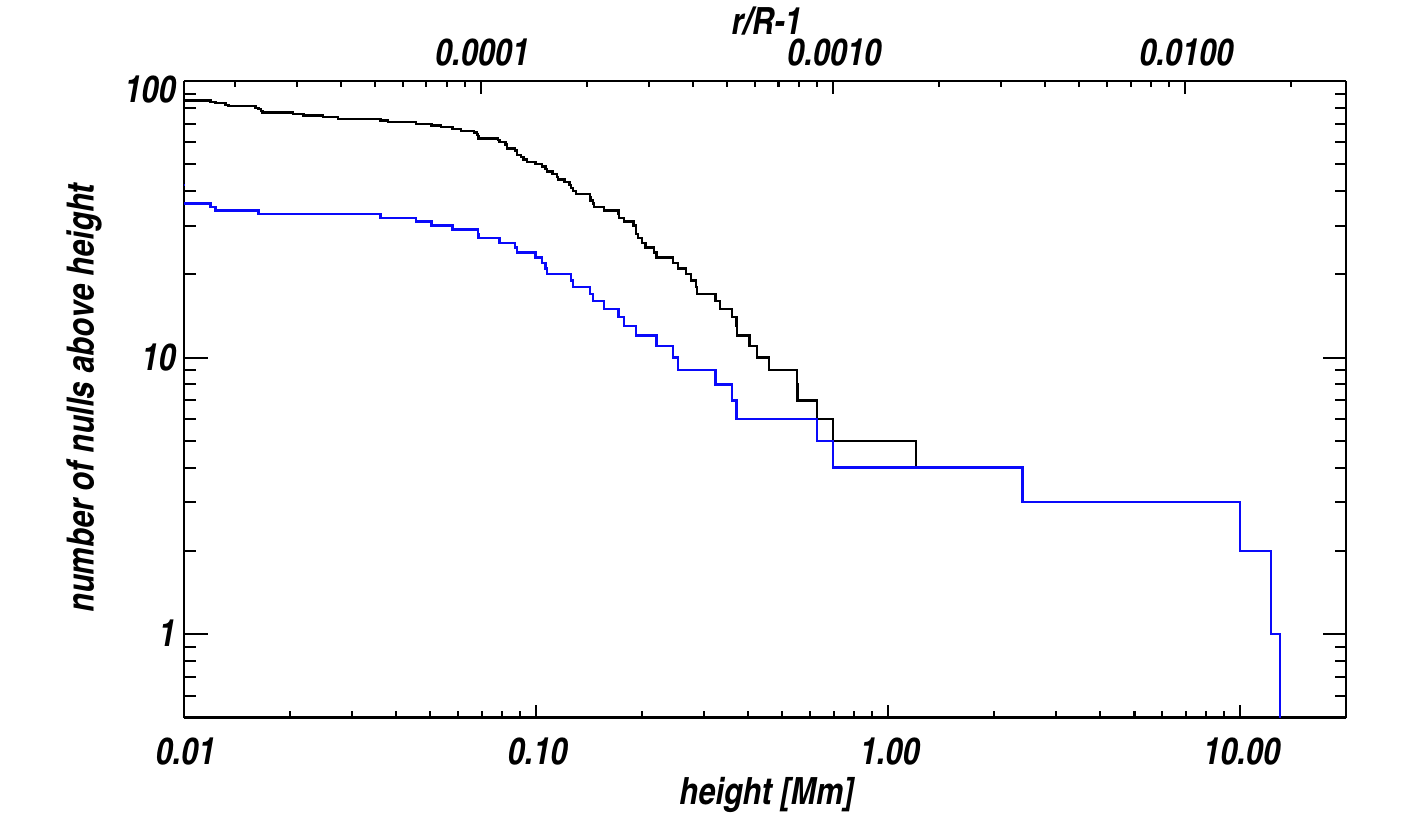}{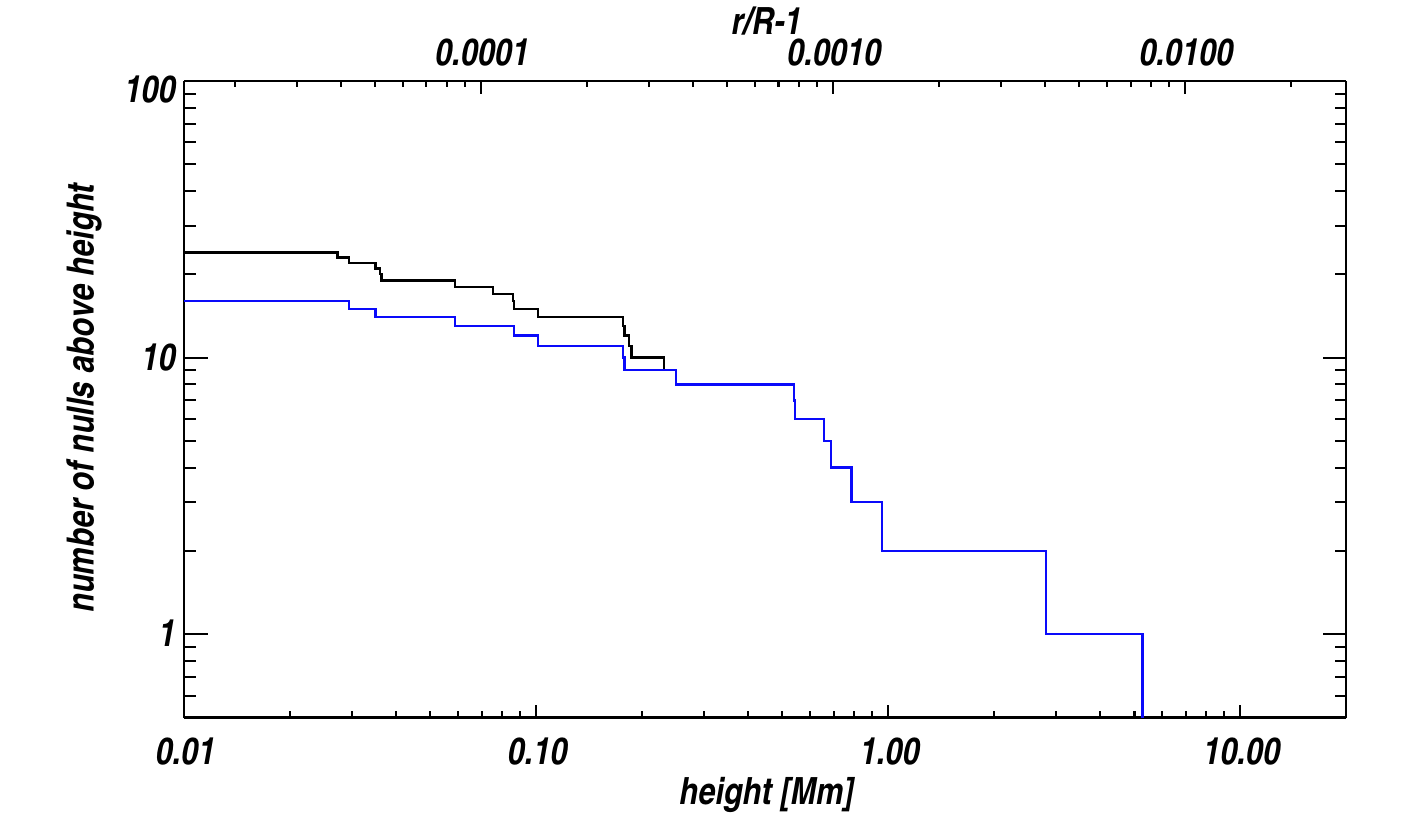}
\caption{Distributions of the height of null points in a PFSS model for NOAA AR\,11112 at 17:48\,TAI on 2010 October 16 (left) and NOAA AR\,11158 at 01:00\,TAI on 2011 February 15 (right).
The black curve shows the total number of nulls points above a given height, while the blue curve shows the total number of nulls that are also present in the PFSS model from an HMI magnetogram taken twelve minutes earlier (AR\,11112) or later (AR\,11158).
For both regions, the highest nulls are present at both times, while the majority of the low nulls are not, suggesting that the existence of the higher nulls is robust to noise in the magnetograms.
\label{fig:null_height}}
\end{figure*}

The smaller number of null points found above AR\,11158 compared with AR\,11112 is likely a result of the larger area covered by strong unipolar field.
The vast majority of the null points found for AR\,11112 lie above weak, mixed polarity field, and do not appear to play any role in the event.
Many of these nulls points are low-lying and likely an artifact of noise, as 
pointed out by \cite{Albright1999,LongcopeParnell2009}.
To test this, for both cases, the nulls at the time that was analyzed were compared to the nulls present in a PFSS model from a time twelve minutes different.
The data between the two times are largely independent, so any null point that is present at both times is unlikely to be an artifact of noise.
Changes in the topology between times could be an artifact of the noise or a result of real evolution of the photospheric field.

Figure~\ref{fig:null_height} shows the height distribution of the null points for the two regions. 
For both regions, the highest nulls (height $\ga 1$\,Mm) are present at both times and are therefore unlikely to be a result of noise, while the majority of the lower nulls are only present at the time analyzed.
Further, all the nulls that are connected by separators that likely play a role in the events (i.e., those shown in Figs.~\ref{fig:AR11112_skeleton}, \ref{fig:AR11158_skeleton}) are also present at both times. 

\section{The Magnetic Skeleton for NOAA AR\,11158}\label{sec:AR11158_topology}

\begin{figure*}
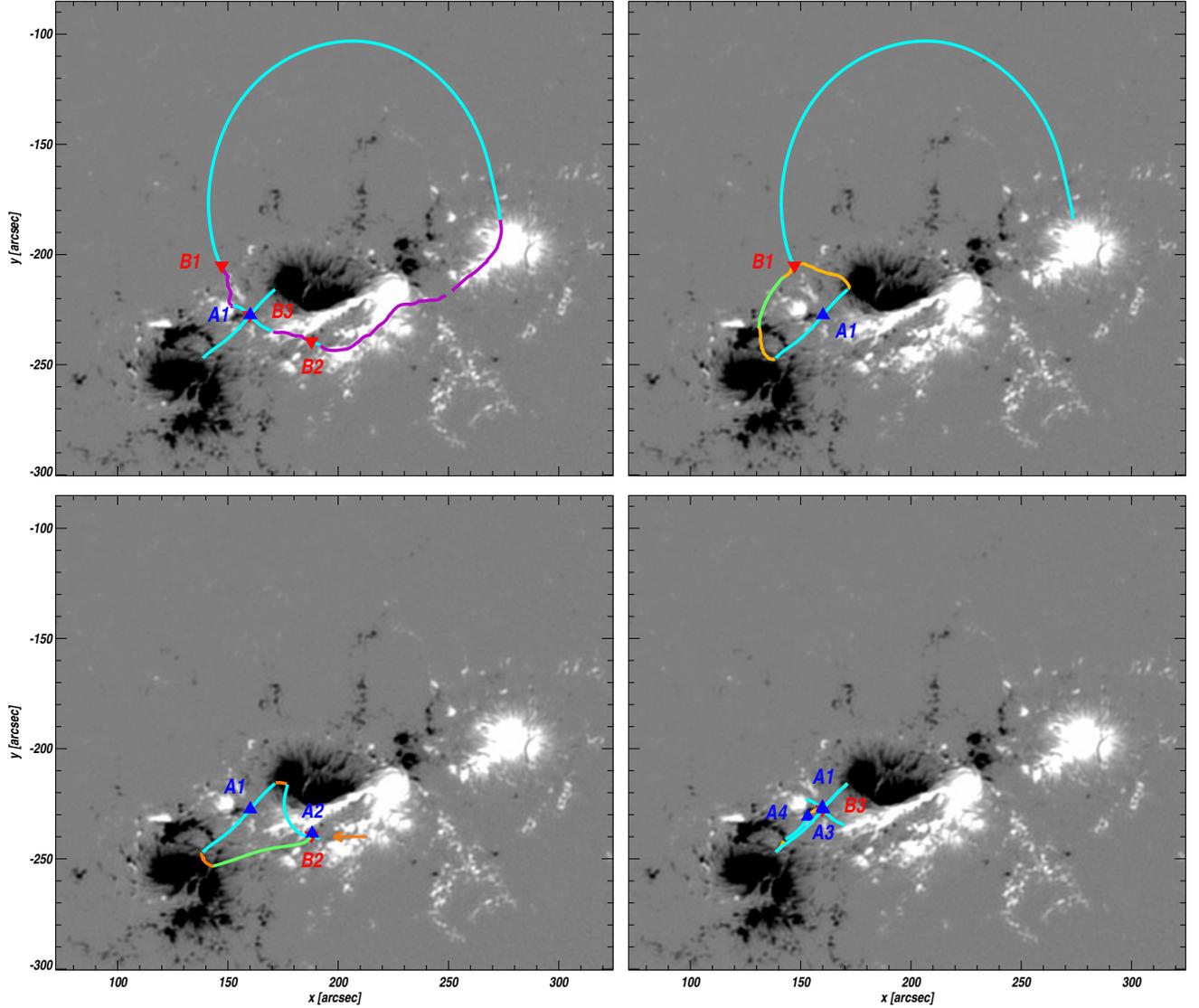

\gridline{\fig{2011-02-15T01-00-00_AR11158_Pr_nls_A1ftr.eps}{0.53\textwidth}{}\hspace{-1.3cm}
          \fig{2011-02-15T01-00-00_AR11158_Pr_nls_B1ftr.eps}{0.53\textwidth}{}}
\vspace{-1.8cm}
\gridline{\fig{2011-02-15T01-00-00_AR11158_Pr_nls_B2ftr.eps}{0.53\textwidth}{}\hspace{-1.3cm}
	  \fig{2011-02-15T01-00-00_AR11158_Pr_nls_B3ftr.eps}{0.53\textwidth}{}}
\vspace{-0.8cm}
\caption{Elements of the magnetic skeleton in a PFSS model for NOAA AR\,11158 at 01:00\,TAI on 2011 February 15 in the same format as Fig.~\ref{fig:AR11112_skeleton}.
The upper left panel shows the spines and fan trace for null \A{1}, along with the spines for nulls \B{1}, \B{2}, and \B{3}, which span gaps in the fan trace.
The upper right panel shows the spines and fan trace for null \B{1}, along with the spines for null \A{1} and a field line in a bald patch separatrix surface (light green), which span gaps in the fan trace.
The lower left panel shows the spines and fan trace for null \B{2}, along with the spines for nulls \A{1} and \A{2} and a field line in a bald patch separatrix surface, which span gaps in the fan trace.
There is a short segment of fan trace, whose location is indicated by the orange arrow, that is obscured by the symbols for the null points.
The lower right panel shows the spines and fan trace for null \B{3}, along with the spines for nulls \A{1}, \A{3}, and \A{4}, which span gaps in the fan trace.
Nulls \A{1}, \B{3}, and \A{3} lie approximately along the line of sight, which is more clearly shown from a different perspective in Figure~\ref{fig:fan_sideview}, right.
The separatrix surfaces for nulls \A{1}, \B{2}, and \B{3} are tunnels with three entrances, while the separatrix surface for null \B{1} is a tunnel with two entrances.
\label{fig:fan_traces}}
\end{figure*}

Section~\ref{sec:AR11158} described the magnetic skeleton of NOAA AR\,11158 in sufficient detail to make comparisons to the flare ribbons. 
Here, we provide additional details for the interested reader wishing to better visualize the skeleton.
Each panel of Figure~\ref{fig:fan_traces} shows the fan trace for one of the null points already shown in Figure~\ref{fig:AR11158_skeleton}, along with only the spines and location of other nulls that are connected by separators (not shown) to this null, and selected field lines in bald patch separatrix surfaces that explain the gaps in the fan traces.

The fan trace for null \A{1} shown in the upper left panel is relatively straightforward, with three disjoint segments, connected by the spine field lines of nulls \B{1}, \B{2}, and \B{3}, making its separatrix surface a tunnel with three openings.
In this view, the segment of the fan trace that is largely obscured in Figure~\ref{fig:AR11158_skeleton} by the separator connecting \A{1} and \B{1} is clearly visible.

The upper right panel shows the fan trace for null \B{1}, which has two disjoint segments, connected at one pair of ends by the spine field lines from null \A{1}, and at the other pair of ends by a field line in a bald patch separatrix surface, making \B{1}'s separatrix surface a tunnel with two openings.

\begin{figure*}
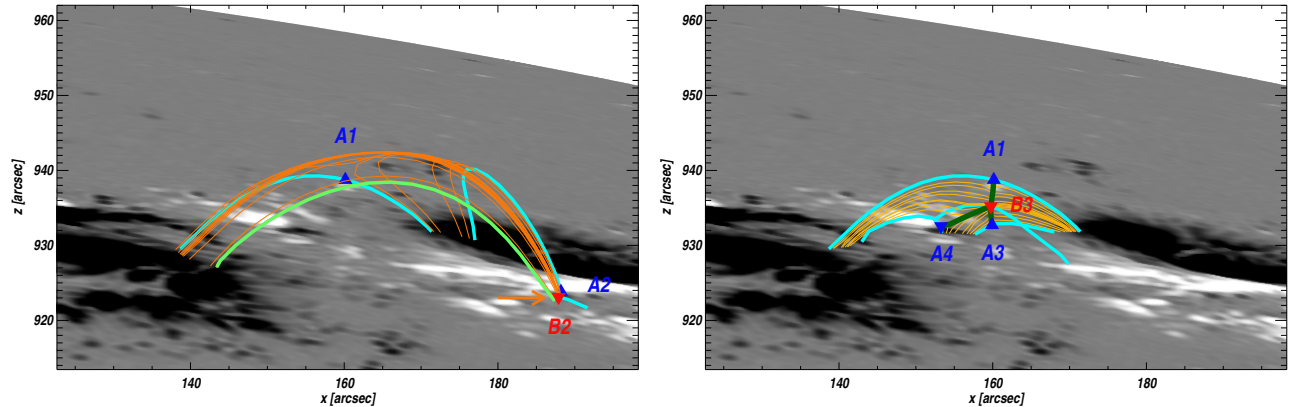

\gridline{\fig{2011-02-15T01-00-00_AR11158_sideview_Pr_nls_B2fan.eps}{0.5\textwidth}{}\hspace{-0.9cm}
	  \fig{2011-02-15T01-00-00_AR11158_sideview_Pr_nls_B3fan.eps}{0.5\textwidth}{}\hspace{-0.0cm}}
\vspace{-0.8cm}
\caption{Field lines in the separatrix surface of null \B{2} (left) and \B{3} (right) in a PFSS model for NOAA AR\,11158 at 01:00\,TAI on 2011 February 15, viewed along the Sun's polar axis from the south.
The separatrix surface for each of these nulls is a tunnel with three openings.
For null \B{2}, the openings follow the spine field lines of nulls \A{1} and \A{2}, plus a bald patch separatrix surface.
The field line in the bald patch separatrix surface at the edge of the null separatrix surface is shown in light green.
There is a short segment of fan trace, indicated by the orange arrow, that is largely obscured by the symbol for null \B{2}.
For null \B{3}, the openings follow the spine field lines of nulls \A{1}, \A{3}, and \A{4}.
\label{fig:fan_sideview}}
\end{figure*}

The lower left panel shows the fan trace for null \B{2}, which has three disjoint segments, connected by the spine field lines of nulls \A{1} and \A{2}, and by a field line in a bald patch separatrix surface, making its separatrix surface a tunnel with three openings.
This is further illustrated in Figure~\ref{fig:fan_sideview}, left.

The fan trace for null \B{3} shown in the lower right panel of Fig.~\ref{fig:fan_traces} also has three disjoint segments, connected by the spine field lines of nulls \A{1}, \A{3}, and \A{4}, more clearly seen in Figure~\ref{fig:fan_sideview}, right as nulls \A{1}, \B{3}, and \A{3} lie approximately along the line of sight. 
The short segment of fan trace between the footpoints of the spine field lines for nulls \A{3} and \A{4} is roughly cospatial with a very small flare ribbon visible in the lower right panel of Figure~\ref{fig:AR11158_skeleton}, but there is no obvious correspondence to the segment of fan trace between the footpoints of the spine field lines for nulls \A{1} and \A{4}, which suggests that the separator connecting nulls \A{1} and \B{3} is not a site for reconnection in this event, and why it is not discussed in \S\ref{sec:AR11158}.

\bibliography{references}{}
\bibliographystyle{aasjournal}

\end{document}